# Characterizing References from Different Disciplines: A Perspective of Citation Content Analysis

Chengzhi Zhang [*], Lifan Liu, Yuzhuo Wang

Department of Information Management, Nanjing University of Science and Technology, Nanjing, 210094, China

**Abstract:** Multidisciplinary cooperation is now common in research since social issues inevitably involve multiple disciplines. In research articles, reference information, especially citation content, is an important representation of communication among different disciplines. Analyzing the distribution characteristics of references from different disciplines in research articles is basic to detecting the sources of referred information and identifying contributions of different disciplines. This work takes articles in PLoS as the data and characterizes the references from different disciplines based on Citation Content Analysis (CCA). First, we download 210,334 full-text articles from PLoS and collect the information of the in-text citations. Then, we identify the discipline of each reference in these academic articles. To characterize the distribution of these references, we analyze three characteristics, namely, the number of citations, the average cited intensity and the average citation length. Finally, we conclude that the distributions of references from different disciplines are significantly different. Although most references come from Natural Science, Humanities and Social Sciences play important roles in the Introduction and Background sections of the articles. Basic disciplines, such as Mathematics, mainly provide research methods in the articles in PLoS. Citations mentioned in the Results and Discussion sections of articles are mainly in-discipline citations, such as citations from Nursing and Medicine in PLoS.

## 1 Introduction

The promotion of extensive multidisciplinary cooperation is the key to coping with the future public information crisis (Xie et al., 2020). Academic paper, especially the reference information in citation contents, is a major media that reflects the communication among scientists and research fields (Zhuge, 2006). References from other disciplines can show the information flow into the research field and conducive to the proposing of new methods to solve complex practical problems (Bammer, 2012; Hassan et al., 2018; Gates et al., 2019).

According to citation information, researchers can detect the sources of information and academic importance of different disciplines, such as Bertin et al. investigated the invariant distribution of references to show the different roles citations have in the scholarly communication process (Bertin et al., 2016). The location of references in papers has long been concerned (Voos and Dagaev, 1976).

Therefore, many scholars seek to better understand the research status and influence of different disciplines based on academic papers and the location of references in them (Zhuge, 2006; Hassan and Haddawy, 2015; Wang and Zhang, 2018). And some use bibliographic information to explore the performance of different disciplines in citing and cited works (Steele and Stier, 2000; Chen et al., 2018; Chen et al., 2019). However, these studies mainly use traditional citation analysis and focus on one discipline. Besides, they may ignore important information combine with location and quantity of references, like characteristics of the distribution of the citations from different disciplines. Therefore, this paper seeks to characterize the references from different disciplines based on CCA (Citation Content Analysis) (Chubin and Moitra, 1975; Zhang, Ding and Milojević, 2013).

Different from previous studies, in this study, we use citation and reference information to explore the distribution of references from different disciplines. We not only consider the number and location of citations, but we also apply CCA to analyze the content of citations. We collect full-text articles in PLoS and use the classification system of Scopus (https://www.scopus.com/), which has 27 first-level disciplines.

To be more specific, we can conclude that the research mainly answered the following two questions:

*RQ1*．Are there any differences in the distribution of the references in different section type from the 27 disciplines in PLoS?

*RQ2*．Do the distributions of the references have significant differences among the 27 disciplines in PLoS according to citation analysis and CCA?

In this research, we answer the questions and summarize the contributions in three respects. First, this study characterizes the references from different disciplines. We use the full-text articles in PLoS and further apply a general research method to compare the distributions of the references from different disciplines.

Second, we identify the disciplines of the journals without knowing the discipline in data processing. In this research, we believe that the discipline of the reference is the same as the journals from which the references come. However, almost 70% of the journals have no specific discipline. Therefore, without data processing, we find that the dataset and conclusions are quite different and unreasonable.

Third, this work explores the distributions of different disciplines when cited and proves that there are significant differences among the distributions from different disciplines. This information is essential for revealing the influences of different disciplines in the papers and important in identifying the contributions of different disciplines.

In the following section, we will investigate related work and introduce the dataset and the process of our work in detail. To determine the values of all characteristics in the distribution of references from different disciplines, we test the differences among the distributions of references from different disciplines and discuss the results. Finally, we conclude our findings in this study and introduce future work.

## 2 Related works

Our work aims to characterize the references from different disciplines using CCA. In this section, we investigate the works related to our research, including a brief review of the distributions of the references from different disciplines and Citation Content Analysis.

**2.1 Distributions of the references from different disciplines**

Research on the distributions of references from different disciplines has long been one main topic in the field of Bibliometrics. In behavioral sciences, citation habits are thought to be influenced by the discipline of reference (Lange, 1985). Radicchi et al. (2008) analyzed the universality of citation distributions based on the full text in a single publication, and they only assessed several disciplines using the numbers of citations (Radicchi, Fortunato and Castellano, 2008). Exploring the distribution of references from disciplines in academic papers can help to better understand the influences of different disciplines in research (Chen et al., 2019).

Steele and Stier (2000) researched Environmental Science through the study of the diversity of the disciplines of citations. They concluded that in many researches, knowledge from other disciplines can have large impacts on these works. Similarly, Rinia et al. (2002) classified 15 fields, including Basic Life Science, Computer Science, Mathematics and others. Recently, Sánchez-Gil et al. (2018) studied the density trends of the references from some major disciplines. More tools for information gathering and processing make it easy to obtain more details on citation information. Park and Kang (2009) gathered the scientific papers cited in Korean patents. They mainly used the citation information to measure the differences between the knowledge transfer patterns between scientific papers and patents in some disciplines (Park and Kang, 2009). Some researchers analyzed the performance of citations in one discipline from other disciplines to demonstrate the impact of this discipline (Hessey and Willett, 2013; Wade et al., 2006). However, they all worked in a specific research field such as LIS (Library and Information Science). Boyack et al. (2018) found that there is a high difference in the age and citation counts of the distributions of references (Boyack et al., 2018).

The research of the locations of citations is a basic work for exploring the distributions of citations. Only by clarifying the position of references can we combine the citation information with their locations. Then, we can explore the distribution characteristics and more possibilities. Bertin et al. (2016) divided the structure of articles into four parts to investigate the invariant distributions of references in scientific articles using bibliometrics (Bertin et al., 2016). Halevi and Moed (2013) researched the distributions of citations. They analyzed the ratios of different disciplines cited and the positions of the citations (Halevi & Moed, 2013). However, their work only focused on the influence of the *Journal of Informetrics* (2007) on other research fields. Thelwall (2019) studied the systematic differences between 22 disciplines in the most cited sections but found no differences (Thelwall, M., 2019). These researches mostly concentrate on specific research areas or use traditional bibliometrics such as citation analysis, and they do not consider both the citation content and location. Different from previous works, we combine citation sentences and their positions in this study. Then, we integrate several characteristics to show the distributions of the references from 27 disciplines and their differences.

**2.2 Citation Content Analysis**

Citation Content Analysis (CCA) is a method for analyzing citation content that combines quantitative and qualitative measurements. Several previous works have stated that we should apply content analysis in citation analysis (Chubin and Moitra, 1975; Mccain and Turner, 1989). The structure of a paper helps to note the locations of citations (Maricic et al., 1998). Different from citation analysis, the CCA method can reveal the content and context of citations. It is important in investigating the contribution of a cited entity or work or other information. However, traditional citation analysis is quantitative (Pinski and Narin, 1976). In some basic researches, it is effective at revealing intercitations or cross citations, demonstrating research priorities, and so on (Hassan and Haddawy, 2015). The problems of traditional citation analysis were proposed in the 20th century. Some researchers proposed that this method is controversial and should be used cautiously because some factors such as different citations have different roles, and their importance may produce errors in the results (Garfield, 1979; MacRoberts and MacRoberts, 1996).

In 1975, Chubin and Moitra researched the content analysis of references and questioned the traditional citation count. They conducted analysis of citations and references separately (Chubin & Moitra, 1975). Though they considered the content of references such as identifying positive or negative citations, their research lacked more full-text content information. In recent years, the OA (Open Access) movement has provided access to a large amount of full-text data from research articles (Eysenbach, 2006). The OA movement allows researchers to extend their research to more research topics and go deeper into the potential information of the content of papers. Nevertheless, CCA is still limited by the dataset and technology.

With the support of natural language processing and many other technologies, more researchers can analyze citation content from different disciplines or other topics. Zhang et al. (2013) considered syntactic and semantic analysis and proposed a framework for CCA (Zhang, Ding and Milojević, 2013). Then, Ding (2014) proposed that content-based citation analysis, which can distinguish the values and weights of different citations, would be the next generation of citation analysis (Ding, Y., et al., 2014). Hassan et al. (2018) used CCA to distinguish the different importances of various citations (Hassan et al., 2018).

In general, the existing works on the distribution of references and Citation Content Analysis have received widespread attention. In addition, the OA movement and new technologies provide researchers with more detailed information from data collection and more approaches for data processing and analysis. However, most previous works did not consider all disciplines, and the methods and results may only be practical in specific fields. Therefore, in this paper, we use the classification system from Scopus and identify the disciplines of the most referenced records from all disciplines, which have generally been ignored by previous researches. Based on this relatively complete dataset, we seek to better investigate the disciplinary distributional patterns of references and use CCA to analyze the characteristics of the distributional patterns of references.

## 3 Methods

In this section, we introduce the main steps in this research, including data processing and the analysis of the characteristics of reference distributions. The framework of the method in this study is shown in Fig. 1.

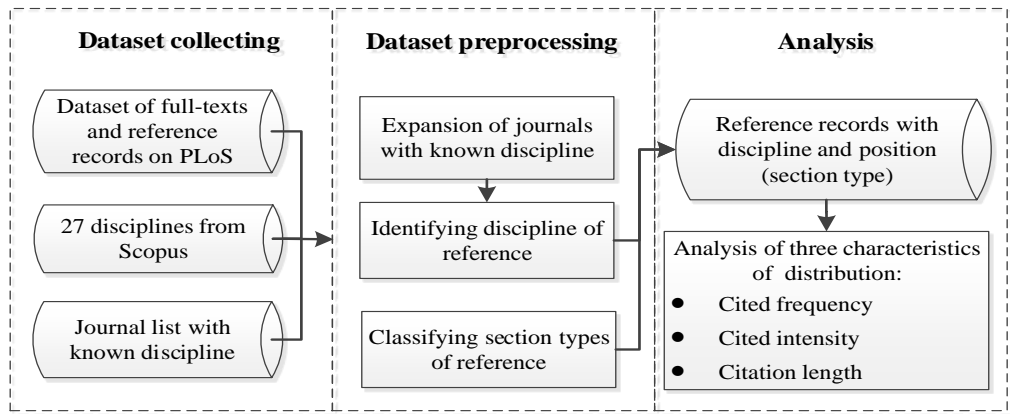

**Fig. 1. Framework of method in this study**

### 3.1 Dataset collecting

PLoS is a typical publisher that provides large number of full-text scientific articles for many researchers (Eysenbach, 2006). There are seven journals in the life science and medicine fields published on PLoS with free access to the full text. We retrieved 210,334 articles in XML format, whose types are 'Research Articles' and published from 2003-10-01 to 2017-12-31, and then downloaded them from PLoS. The numbers of articles we downloaded from seven journals are shown in Table 1.

**Table 1. Seven journals published on PLoS**

| Journals | Abbreviation | # Download articles | Number of articles we choose |
|---|---|---|---|
| PLOS ONE | PONE | 184,543 | 132,228 |
| PLOS GENETICS | PGEN | 6,426 | 6,236 |
| PLOS PATHOGENS | PPAT | 5,494 | 5,381 |
| PLOS NEGLECTED TROPICAL DISEASES | PNTD | 5,004 | 3,403 |
| PLOS COMPUTATIONAL BIOLOGY | PCBI | 4,958 | 4,028 |
| PLOS BIOLOGY | PBIO | 2,411 | 2,032 |
| PLOS MEDICINE | PMED | 1,454 | 799 |
| PLOS CLINICAL TRIALS | PCTR | 44 | 27 |
| TOTAL | | 210,334 | 187,396 |

Note: PLOS Clinical Trials was merged in August 2007 with PLOS ONE.

We extract the DOI, title, heading of first-level sections, references and citation content (sentences with a marked reference (Chubin & Moitra, 1975)) and record it into the database by parsing the articles. It should be noted that there are series of types of reference sources that appear in all articles, including 'journal', 'book', and 'other', and even many other sources with spelling errors (e.g., "journel"). We classified all forms and found that 'journal' had the largest proportion of all references. Finally, we maintain 187,397 articles where the proportion of 'journal' comprises more than 80% of all references records and just focus on them.

**3.2 Dataset preprocessing**

Data processing, as shown in Fig. 2, has two main steps. First, we need to identify the disciplines of the references, and this is the essential work to explore the characteristics of references from different disciplines. Second, to analyze the distribution of reference, we must find the position (section type) where the reference has been cited.

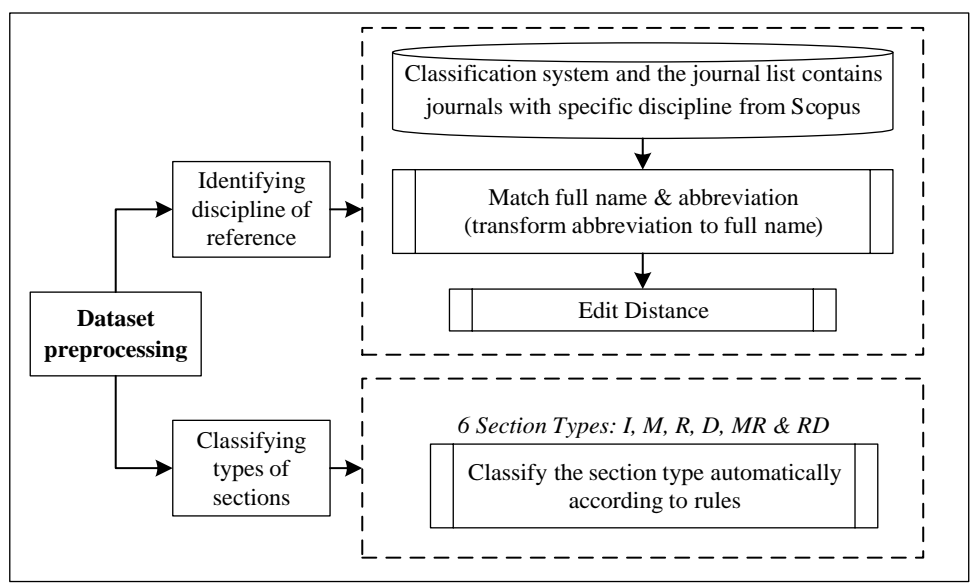

Fig. 2. Framework of data processing

*3.2.1 Identifying discipline of references*

Scopus has a fine-grained classification system of disciplines and provides a journal list that contains more journals with specific disciplines. The classification system of Scopus contains 27 first-level disciplines, including Medicine, Multidisciplinary, Mathematics, Social and others; and the journal list covers 23,359 journals[1]. For example, *Nature* is multidisciplinary, *J Biol Chem* (*The Journal of Biological Chemistry*) belongs to Biochem Mol Biol (Biochemistry Genetics and Molecular Biology), and *Lancet* belongs to Medicine. We use the first word of the discipline name to represent the corresponding name, as shown in appendix A.

In this study, we get a total of 9,528,160 records of references that are 'journals'. After removing duplicates and normalizing the same journal names, we obtained 409,728 journal titles, including the full name and various abbreviations (e.g., the full name is *Proceedings of the National Academy of Sciences of the United States of America*, and its abbreviations include *PNAS*, *proc natl acad sci*, *proc natl acad sci usa*, *proc. natl. acad. sci. usa* and so on). Then, we identify the disciplines of the references according to the name of the journals. The discrimination process of the disciplines of references is mainly divided into the following parts:

(1) Match journal names to the journal list directly

First, we directly match all the records of reference to the journal list from Scopus. Only 30.48% of the records (2,904,093) can be matched to their disciplines. This is because a large number of the journal names from the records are abbreviations. Therefore, we need to associate the abbreviations of the journal name with its full name.

---

[1] The date when all journals supporting by Scopus were counted was 2018/09/18.

(2) Match full names and abbreviations of journals

Following the first step, we obtain the full name of abbreviations through the journal title list from the NLM (National Library of Medicine, https://www.nlm.nih.gov/). Then, we identify the disciplines of references by matching the full name of journals in the references with the journal list in Scopus, and 71.90% of references (6,851,139 records) are matched with their disciplines.

(3) Match journal disciplines using the Edit Distance algorithm

For the remaining journals with unknown disciplines, we call them journal U. The edit distance algorithm is utilized to calculate the steps of the edits (e.g., delete and insert) that can convert the names of journals from journal U into the name of journals with known disciplines (Marzal & Vidal, 1993). The discipline of a journal that has the minimum number of steps is the discipline of journal U. After manual checking, 92.66% of the reference records (8,828,654) have a discipline.

After identifying the disciplines of references, we selected 154,134 articles from the dataset. In each article, the reference records with certain disciplines can cover no less than 80% of all records. Meanwhile, in order to ensure the validity of the data, we maintain the articles where the proportion of 'journal' references comprised more than 80% of the article's total reference records.

*3.2.2 Classifying types of sections*

Citations are not equal since their importance can also depend on their positions in the content of article in addition to their frequency (Voos and Dagaev, 1976). In this study, we classify the first-level sections into six section types, namely, the IMRaD structure, MR and RD (Bertin & Atanassova, 2018). In IMRaD, I means Introduction, Overview, or Background.. Similarly, M represents Methods or Materials. R represents Results or Findings, and D represents Discussion or Conclusion. If the section title has both M and R, i.e., Methods and Results, we classify this section type as MR; and similarly, we classify Results and Discussion as RD.

We classify the first-level sections into different categories according to the heading of each first-level section. In this paper, we obtain 733,302 headings of the first-level section from 154,134 PLoS articles. After the processes of converting all headings to lowercase and removing the numbers from headings, we then remove the high-frequency invalid headings such as *funding supporting information, acknowledgments, appendix*, and so on. Using the headings of sections and the body text of articles, we examine the most common section names and use them to identify all first-level sections in articles into six types automatically. Finally, 99.4% of these sections (625,914) have exact types. As Fig. 3 shows, the distributions of the four section types including I, M, R, and D are similar, but there are few citations that belong to section MR and section RD. This shows that the number of citations in section MR in the PLoS dataset is less, and it is related to the author's writing habits. Scholars prefer to describe the results and discussions separately. Therefore, the number of citations in section MR is very low.

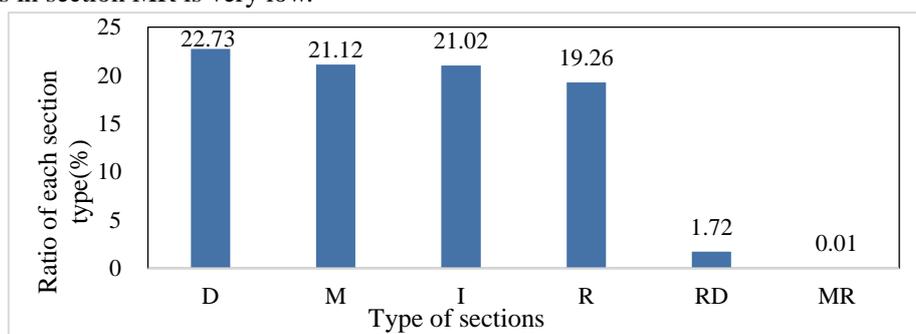

**Fig. 3. Proportion of different type of sections**

## 3.3 Characterizing references from different disciplines based on CCA

There are three characteristics we use to show the differences among disciplines quantitatively in this paper: the number of citations, the average cited intensity and the average citation length.

(1) Number of citations

**Definition 1**: We use $N$ to represent the number of citations.

We use $N_{ij}$ to denote the number of discipline $i$ cited in section $j$ and $N_i$ to denote the number of discipline $i$ cited in all articles ($i = 1, 2, \ldots\ldots 27; j = I, M, R, MR, D, RD$). $i$ is the ID of the discipline shown in Appendix A, and $j$ is the section type (e.g., $N_{1M}$ represents the number that references from Medicine appear in section M).

**Definition 2**: The proportion of citations is the ratio at which the number citations from one discipline appears in all citation records. In this study, we use $P_{ij}$ to represent the percent of citations in total.

$$P_{ij} = N_{ij}/N_i \tag{1}$$

In this paper, $P_{ij}$ is used to compare the distribution of 27 disciplines in different section types, where $\sum_j P_{ij} = 1$.

(2) Average cited intensity of reference from different disciplines

**Definition 3**: The average cited intensity represents the average number of citations of a reference from one discipline mentioned in one paper. For one reference in an article, the cited intensity equals the number of times that it appears in this article. For discipline $i$ in an article, if the number of references from discipline $i$ is A and the references appear B times in this article, B/A is the average cited intensity of this discipline in this article. Therefore, for discipline $i$ in all articles, where $n_i$ is the number of references, and $N_i$ is the number of references that appear, we can obtain the average cited intensity as follows.

$$ACI_i = N_i/n_i \tag{2}$$

We also count the number of references from discipline $i$ in section $j$ and record it as $n_{ij}$. According to the number of citations $N_{ij}$, we can obtain the average intensity of discipline i in section $j$ as follows.

$$ACI_{ij} = N_{ij}/n_{ij} \tag{3}$$

(3) Average citation length of the citation content from different disciplines

**Definition 4**: The citation length is the number of words of citation content (Wang, Y. and Zhang, C., 2019). We use $ACL_i$ (average citation length of discipline $i$) to represent the average number of words of all citations from discipline $i$.

$$ACL_i = L_i/N_i \tag{4}$$

$L_i$ is the total number of words of all citation content from discipline $i$, and $N_i$ is the number of citations of discipline $i$, as mentioned before.

Regarding different locations, we also obtain the total number of citation content words in different section types. For section j, we use $L_{ij}$ to represent the total number of words of citation content and $N_{ij}$ to represent the number of citations from discipline $i$. Similar to the average cited intensity, we use $ACL_{ij}$ to represent the average citation length of citation content in section $j$ from discipline $i$ according to formula (5) as follows.

$$ACL_{ij} = L_{ij}/N_{ij} \tag{5}$$

# 4 Results

This section shows all three characteristics of the distribution and answers the questions we raised in this research. We obtain $N_i$ and $N_{ij}$ (the numbers of citations of each discipline and in different section types, respectively), and $P_i$ and $P_{ij}$ (the percent of $N_i$ and $N_{ij}$ in total) using traditional citation analysis. The average cited intensity and average citation length are calculated by CCA.

## 4.1 Number of references of different disciplines

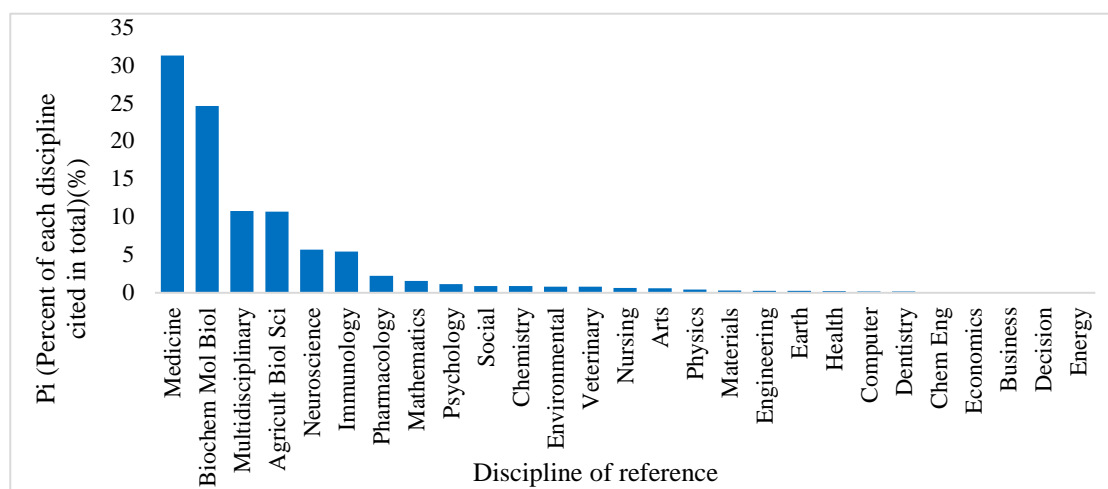

**Fig. 4. Proportion of number of references in 27 disciplines**

In this section, we use traditional citation analysis to obtain the number of references from different disciplines. We first calculated $N_i$ and $P_i$ from 154,134 articles and then used Fig. 4 to illustrate $P_i$, which is the percent of each discipline cited in total.

The top 10 disciplines in Fig. 4 include Medicine, Biochem Mol Biol, Multidisciplinary, Agricult Biol Sci, Neuroscience, Immunology, Pharmacology, Mathematics, Psychology and Social.

**Medicine** We conclude that the top 10 disciplines account for 90.84% of all citation records, and most of the top disciplines belong to the field of Natural Sciences. In the top 10 disciplines, Medicine has the highest number of citations, since most research articles provided on PLoS focus on Medicine, Biochemistry Genetics and Molecular Biology.

**Multidisciplinary** Multidisciplinary ranks second, which indicates that except for Medicine references, the Multidisciplinary references are the most likely to be cited in the PLoS. This result shows that in the fields of life science and medicine, the phenomenon of interdisciplinary research is obvious.

**Mathematics & Social** Moreover, Mathematics ranks eighth among the top 10 disciplines, and Social ranks tenth, different from other topic-related disciplines. In addition, we find that several journals account for many references, such as *Bioinformatics* (Mathematics), *Development* (Social) and others. This shows that basic disciplines such as Mathematics also provide a large proportion of citations, and Social is also important in many research fields such as Medicine.

In addition to the top ten disciplines, Arts and Computer have lower proportions, and the other three disciplines, Business, Decision and Energy, occupy a very small proportion of citations at approximately 0.05%. Therefore, there are fewer scientific articles related to these disciplines in our dataset.

## 4.2 Distribution of the number of citations of 27 disciplines in different section types

*4.2.1 Number of citations in different kinds of section*

The distribution of citations can be generated based on the number of citations in different sections. With the section types where citations are located, we can obtain $N_j$ and $P_j$ (the number of citations and the percent of citations in different kinds of section, respectively) related to the position information. The distributions of citations in the six section types are shown in Fig. 5.

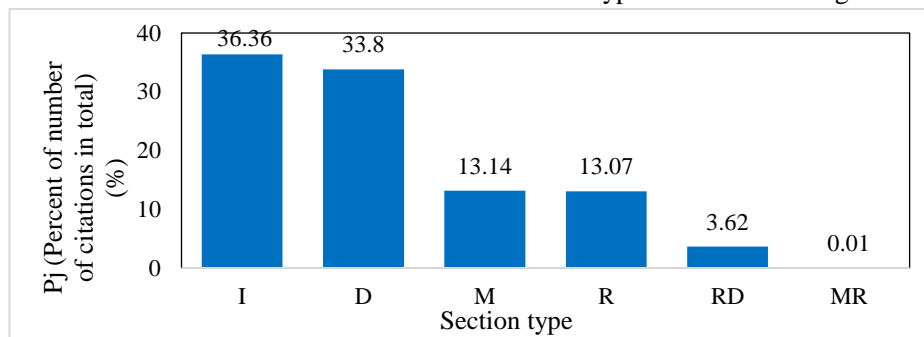

**Fig. 5. Proportion of number of citations in 6 section types**

References are mainly cited in sections I and D. Section I has the highest ratio of citations while section MR has the lowest. In Fig. 5, section I and section D have similar proportions, as do section M and section R. Section RD has a small percentage, and the percentage of section MR is approximately 0.01%. Therefore, these results show that most of the published articles on PLoS are laid out in a fixed structure such as IRMaD, and fewer researchers tend to combine the Methods and Results in the articles on PLoS.

*4.2.2 Proportion of the number of citations of each discipline in different section types*

**We answer RQ1 and RQ2 in this section.** The results show that one of the characteristics $N_{ij}$ (number of citations in different section types) reflects differences in the distributions of references from different disciplines.

To understand the distributions in different section types of the 27 disciplines, we calculate all $P_{ij}$s (percent of citations of each section type for each discipline in total). For each discipline, we use $P_{ij}$ to represent the distribution of each respective discipline. In order to better represent the difference among disciplines, the results are illustrated in Fig. 6.

The results shown in Figure 6 demonstrate that the distributions of 27 disciplines are different. We analyze the results in conjunction with Figs. 4 and 6 to explore the differences of the distributions of the 27 disciplines. We see that Medicine has the highest number of citations. In addition, many other disciplines, such as Biochem Mol Biol, Multidisciplinary, Agricult Biol Sci, Neuroscience and Immunology, have similar distributions to the distribution of Medicine. However, some disciplines are different. Mathematics is the eighth highest discipline in Fig. 4 and more likely to appear in section M. Decision and Computer are similar to Mathematics regarding their distributions of citations, indicating that these basic disciplines are all mostly cited in section M and less cited in other section types. Social ranks tenth in Fig. 4 and is most cited in section I, but it also has the highest proportion among all disciplines in section R. As for Energy, Fig. 4 shows that it has the lowest proportion of citations, which proves that the discipline is the least relevant to the research area of the dataset. However, in Fig. 6, we find that the proportion in one section accounts for 58.57% of the total number of citations. The reason for this mainly comes from the small size of the Energy dataset.

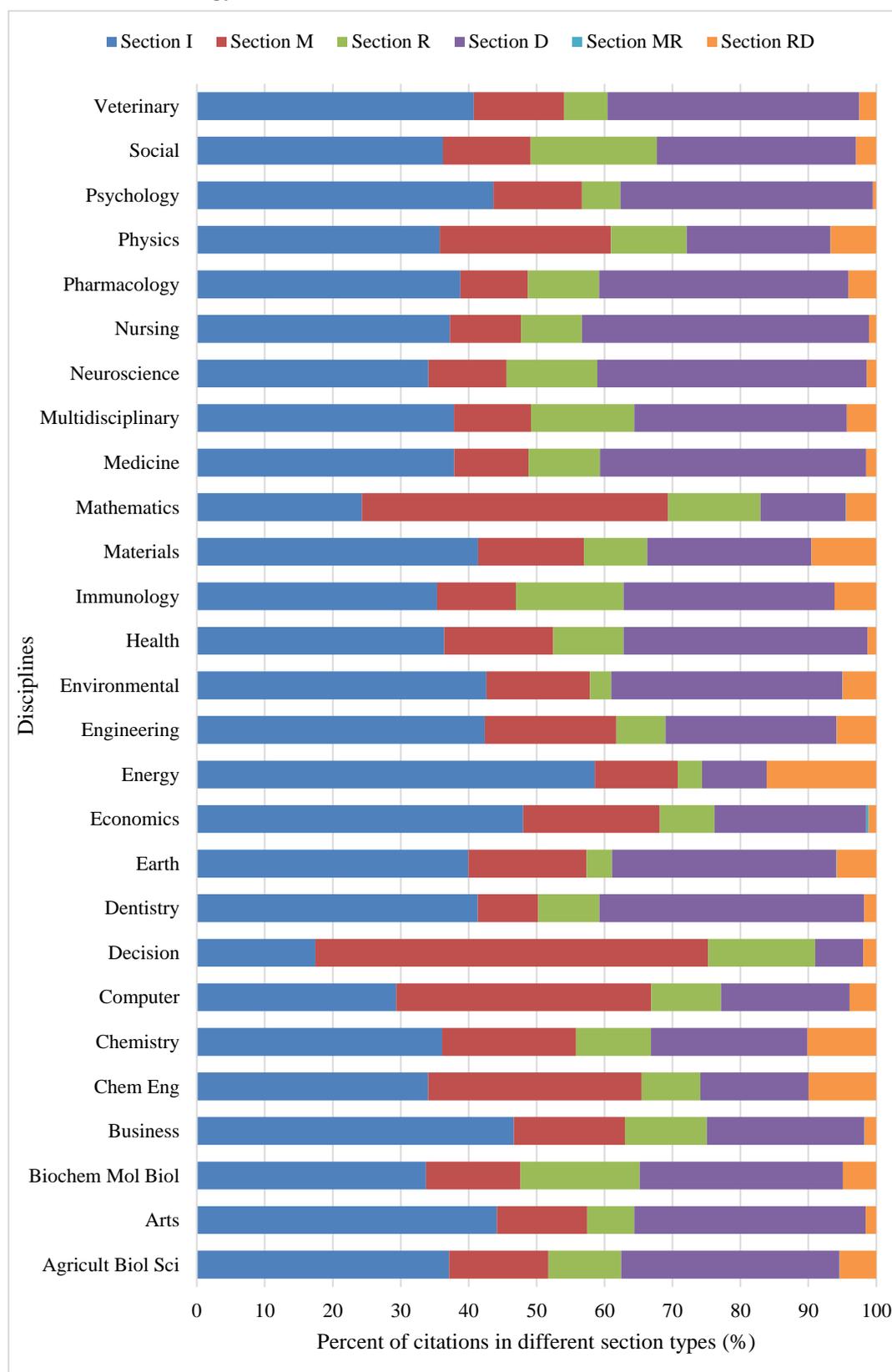

**Fig. 6. Percent of citations in six section types of 27 disciplines**

Obviously, for all disciplines, the number of citations varies among different sections. We use ANOVA (Analysis of Variance) to test if there are significant differences between the proportion of 27 disciplines in each section type. The proportions in section MR and RD are small, so we just take the results of four sections into consideration. The null hypothesis is that the variances of different section types are the same. The results are shown in Table 2.

**Table 2. ANOVA Test of percent of citations in different sections.**

(a) Standard deviations of the proportion of citations in 4 section types.

| Section | Mean | Standard Deviations | Standard Error |
|---|---|---|---|
| I | 0.3819 | 0.0760 | 0.0146 |
| M | 0.1870 | 0.1158 | 0.0223 |
| R | 0.1026 | 0.0412 | 0.0079 |
| D | 0.2832 | 0.0965 | 0.0186 |
| Total | 0.2387 | 0.1355 | 0.0130 |

(b) Test of Homogeneity of Variances.

| Proportion | Levene Statistic | df | Significance |
|---|---|---|---|
| Based on Median | 2.462 | 3 | 0.067 |
| Based on Median and with adjusted df | 2.462 | 3 | 0.071 |

(c) ANOVA test of distribution of citations in 4 section types.

| Source of difference | Sum of Squares | df | Mean Square | F | Significance |
|---|---|---|---|---|---|
| Between Groups | 2.8857 | 3 | 0.393 | 52.093 | 0.000 |
| Within Groups | 0.8188 | 104 | 0.008 | | |
| Total | 3.7045 | 107 | | | |

(d) Multiple Comparisons of distribution of citations in 4 section types.

| (I) section type | (J) section type | Mean Difference (I-J) | Standard Error | Significance |
|---|---|---|---|---|
| I | M | 0.1949* | 0.0236 | 0.000 |
| I | R | 0.2793* | 0.0236 | 0.000 |
| I | D | 0.0987* | 0.0236 | 0.000 |
| M | R | 0.0843* | 0.0236 | 0.001 |
| M | D | -0.0962* | 0.0236 | 0.000 |
| R | D | -0.1805* | 0.0236 | 0.000 |

Note: *. The mean difference is significant at the 0.05 level.

According to the results shown in Table 2(b), the results of significance in Test of Homogeneity of Variances are higher than 0.05, so we can believe that the test result of ANOVA is meaningful. The values of significance in Tables(c) and (d) are all lower than 0.05, so the null hypothesis is rejected. The result demonstrates that there are significant differences in the distributions of the citations of the 27 disciplines in 4 section types. Therefore, whether from the practical situation or the test results (Kirk, 1996), the difference between the proportion of the 27 disciplines is significant in these sections.

**Section I**: Section I has more than half of the citations of Energy (58.57%). And according to Fig. 3, we already know that Energy has the least number of citations in our dataset. Besides, we also find that the proportions of citations in section I of Economics, Business, Arts, Psychology, Environmental, Engineering, and Materials are all more than 40%.

**Section M**: In section M, the difference among the 27 disciplines is significant since there are only two disciplines in which the proportion is more than 40%. Decision (57.79%) is the most prominent. The other discipline in section M with a value greater than 40% is Mathematics. Both of these disciplines are basic disciplines. In addition, Computer is the other most cited discipline in section M.

**Section R**: Social (18.59%) is mostly cited in section R and Nursing (42.23%) is mostly cited in section D. In section R, even though the Biochem Mol Biol and Immunology are more relevant to the topics of most articles, the proportion of Social in this section is higher than the proportions of these two disciplines. The results show that Social is important in section R, and some other disciplines relevant to the fields of the PLoS are also higher more in this section.

**Section D**: In section D, some disciplines, including Nursing and Medicine obtain higher values, which also indicates that the disciplines cited in section D are more relevant to the topics in PLoS. Except for these disciplines relevant to the fields of the dataset, Arts (34.05%) are mostly cited in section D.

**Section MR & Section RD**: Economics, Econometrics and Finance (0.36%) and Energy (16.15%) are mostly cited in the remaining sections MR and section RD, and the proportions are both small. In addition, we also find that the disciplines cited in section RD have a high probability of being the same as the disciplines cited in section D.

**4.3 Average cited intensity of the 27 disciplines in full text and in different section types**

**We answer RQ1 and RQ2 in this section based on the analysis of the average cited intensity, which is the second characteristic.**

In this research, we consider $ACI_{ij}$ and $ACI_i$ (the average cited intensity of each discipline in different sections and in the full text, respectively). In order to facilitate the comparison of the distributions of different disciplines in different sections, we first calculated the overall average cited intensity (1.29) of all subjects in the full text. We then calculate the average cited intensity of different disciplines in each section. Finally, we obtain the values of $ACI_{ij}$ and $ACI_i$, as shown in Appendix B.

The top 10 values of $ACI_{ij}$ are shown in Fig. 7. We find that the top 10 values are all from sections R and MR. Furthermore, the values of Health (2.17) and Nursing (2.13) in section R are both higher than 2. The top 10 disciplines are most relevant to the fields of the PLoS, such as Dentistry and Medicine. This indicates that there are more in-discipline citations in section R.

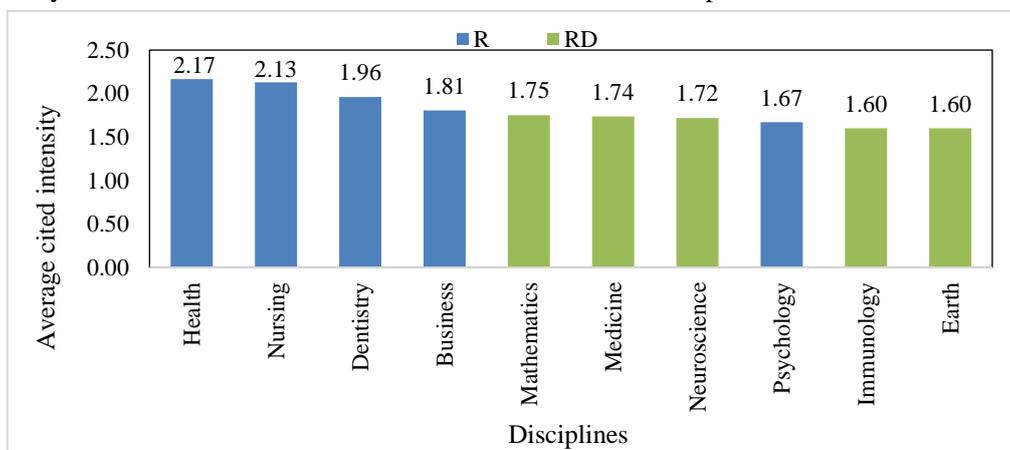

**Fig. 7. Top 10 disciplines of different section types in average cited intensity**

The $ACI_i$ we obtained in our study is similar to the scope of previous research results (Oppenheim, C., Renn, S. P., 1978). According to the $ACI_i$ shown in Appendix B, we find that the highest value is 1.36 (Health) and the lowest is 1.19 (Energy), but the numbers of citations of those two disciplines are both very small. The most likely reason for this result is the small size of the dataset of both disciplines. The standard deviations in Appendix B show that the distributions of $ACL$ in 27 disciplines. Similarly, we use ANOVA to test if there are significant differences in the distributions of average cited intensity from 27 disciplines. The test results are shown in Table 3.

**Table 3. ANOVA Test of average cited intensity from 27 disciplines.**

(a) Test of Homogeneity of Variances.

| Average cited intensity | Levene Statistic | df | Significance |
|---|---|---|---|
| Based on Mean | 1.337 | 26 | 0.147 |
| Based on Median | 0.657 | 26 | 0.894 |

(b) ANOVA test of distribution of average cited intensity.

| Source of difference | Sum of Squares | df | Mean Square | F | Significance |
|---|---|---|---|---|---|
| Between Groups | 0.716 | 26 | 0.028 | 0.872 | 0.646 |
| Within Groups | 4.140 | 131 | 0.032 | - | - |
| Total | 4.856 | 157 | - | - | - |

The value of significance in Table 3(b) is higher than 0.05, which means that $ACI_i$ has no significant difference. Furthermore, the standard deviations shown in Appendix B are low. Although the value in section MR is different and its standard deviation is the highest, the result is influenced by the small dataset, so there is no practical significance.

Therefore, this work shows that the second characteristic that there is no significant difference in the distributions of the average cited intensity of different disciplines.

### 4.4 Average citation length of 27 disciplines in the full text and in different section types

In this part, we also obtain $ACL_{ij}$ and $ACL_i$ (average citation length in different sections and in the full text, respectively). **We answer RQ1 and RQ2 in this section based on the analysis of the average citation length.** The overall average citation length of all subjects in the full text is 30.93 words. Then, we compare the distributions in different sections. We calculate the average citation lengths and the standard deviations of different disciplines in each section. The values we obtain are shown in Appendix C.

As $ACL_{ij}$ and $ACL_i$ shown in Appendix C, the longest average citation content has 32.51 words (Energy), and the shortest has 29.69 words (Chem Eng). It is obvious that most disciplines in which the proportion of citations is lower than 1% have average citation lengths longer than the overall average citation length. The reason for this phenomenon may be that when the number of citations is small, taking the average value will have some contingency, and thus the standard deviations of section MR are extremely high in Appendix C. It also shows that the results obtained by the full text of big data are somewhat explanatory. When a discipline is cited more, its average citation length will be more evenly distributed in different sections. Similar to the method above, we use ANOVA to test if there are significant differences in the distributions of average citation length from 27 disciplines. The test results are shown in Table 4.

The result shown in Table 4(b) is higher than 0.05, which means that there is no significant difference in the average citation lengths of the 27 disciplines. In addition, the difference between the longest citation length and the shortest citation length is less than 5 words, which also indicates that there is no practical significance. This work shows that there is no significant difference in the distribution of the average citation length according to the CCA.

Table 4. ANOVA Test of average citation length from 27 disciplines.

(a) Test of Homogeneity of Variances.

| Average citation length | Levene Statistic | df | Significance |
|---|---|---|---|
| Based on Mean | 1.896 | 26 | 0.010 |
| Based on Median | 0.552 | 26 | 0.960 |

(b) ANOVA test of distribution of average citation length.

| Source of difference | Sum of Squares | df | Mean Square | F | Significance |
|---|---|---|---|---|---|
| Between Groups | 350.119 | 26 | 13.466 | 1.376 | 0.125 |
| Within Groups | 1282.433 | 131 | 9.790 | - | - |
| Total | 1632.552 | 157 | - | - | - |

## 5 Discussion

**5.1 Differences among the distributions of the references from 27 disciplines**

In this research, we explore the differences among the distributions of references in three respects: the number of citations, the average cited intensity and the average citation length. We find something interesting about the characteristics of the number of citations.

Based on the full-text database, the results show that more citations appear in sections I and D (Figs. 4 and 5). This is similar to the work of Thelwall, which shows that the introduction has the most citations in 11 disciplines and the Introduction and Discussion sections contain substantially more citations (Thelwall, M., 2019).

Our results also show the distributions of references from different disciplines (Fig. 6). Interestingly, we find that citations from different disciplines distributed differently. There are some disciplines with unbalanced distributions, such as Decision and Social, which are also the most cited in sections M and R, respectively. Agricult Biol Sci has the highest ratio of the number of citations in section D, 35.19% higher than Veterinary. This is because researchers often compare their results with similar research in the discussion section. Therefore, the in-discipline citations are more likely to appear in section D.

There are significant differences in the distributions of the 27 disciplines, which indicates that the disciplines have different roles in different section types. Section I has the most citations in many disciplines. In section M, basic disciplines, such as Mathematics and Decision, provide most references for PLoS articles. In particular, Social have an important role in section R, and some other disciplines relevant to the fields of the PLoS also give more support in this section. However, in section D, the disciplines relevant to a research topic about PLoS obtain higher ratios, which also show that the disciplines cited in section D are more relevant to the topic of citing papers.

Regarding the average cited intensity and average citation length, the results show no significant differences. Ding et al. counted the frequency of each reference cited in papers, and their results show that the numbers of citations of highly cited references are different among sections (Ding et al.,2013). Different from their work, our results show that the average cited intensity of all references, which is similar to the number of citations, has no significant difference in different sections based on the data from PLoS. In addition, in some works, the results showed that the average number of general papers mentioned increased to 1.50-1.79, and the number of highly cited papers was 1.9 (Wan and Fang, 2014), which is higher than our results. This may be caused by the volume of data, and Ding et al. only considered the highly cited references. Therefore, the differences in cited intensity in each section will be less when considering all reference records, and the size will be larger.

**5.2 Comparison of the distributions of references after and before data processing**

*5.2.1 Proportion of citations of references from different disciplines*

In addition to the results mentioned above, we highlight our main work of expanding the list of journals and identifying the disciplines of the journals and references in data processing. We compare the results before and after this work. As shown in Fig. 8, we first analyze the top 10 disciplines for the total number of citations. The first group in green shows the distribution of the top 10 disciplines before data processing, and the other in blue is the distribution of these 10 disciplines after data processing.

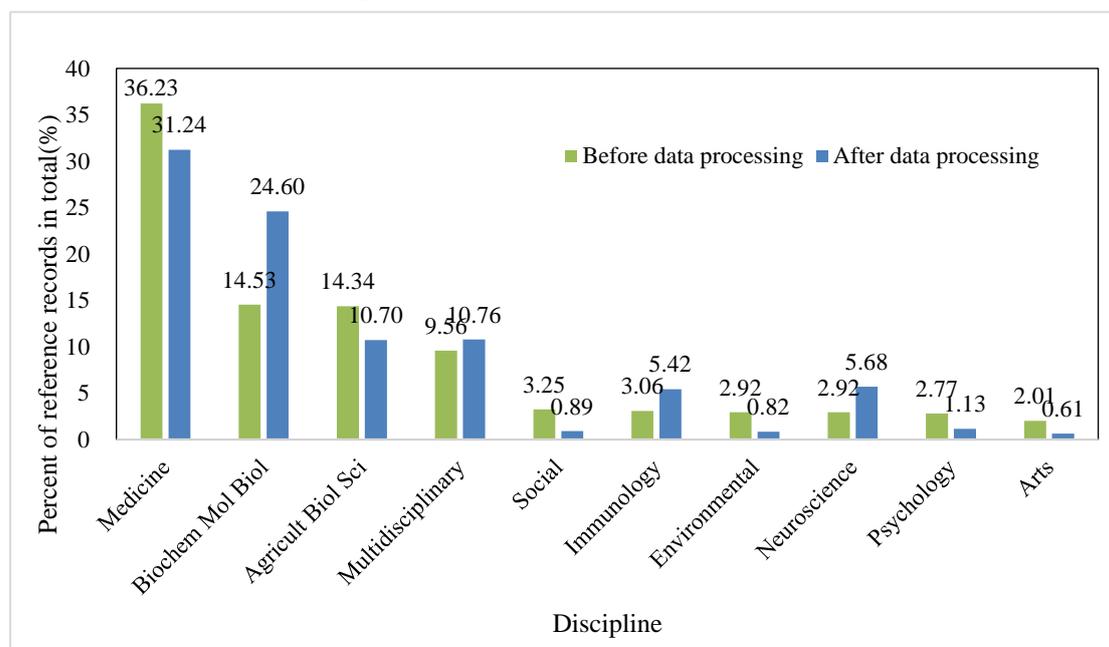

**Fig. 8. Percent of the citations in total before and after the expansion of journal list (top 10 disciplines before the expansion of journal list)**

In Figs. 4 and 8, the top 4 disciplines of the references before and after the journal expansion are the same and the most common disciplines of the citations are both Medicine, but the top 10s are different. Disciplines including Social, Environmental, Psychology and Arts are not in the top 10 disciplines after the expansion of the journal list. This means that the number of citations of disciplines without these four is increased more after identifying the disciplines of the references. The ratio of citations of the top 4 disciplines is almost all higher than 10% and much larger than that of the other disciplines. The ratios of Biochem Mol Biol and Multidisciplinary increased after expanding the journal list. This indicates that most of the extra expansion comes from these disciplines. Besides, Social, Environmental, Psychology and Arts also contribute more to the expansion.

We also find that the data are more unbalanced before the expansion of the journal list. Medicine occupies more of the dataset and Biochem Mol Biol are second, but the number of citations of Biochem Mol Biol is less than half of that of Medicine. The step of identifying the disciplines of the journal and reference is essential and improves data validity. Without the discipline of the citations, we cannot explore the characteristics from different disciplines accurately.

*5.2.2 Distribution of the number of citations in different section types*

**Table 5. percent of citations in six section types of 27 disciplines before the expansion of journal list (comparison after and before data processing).**

| Discipline | Section I (%) | | Section M (%) | | Section R (%) | | Section D (%) | | Section MR (%) | Section RD (%) | Total (%) |
|---|---|---|---|---|---|---|---|---|---|---|---|
| State(A/B) | A | B | A | B | A | B | A | B | A | A | |
| Agricult Biol Sci | 37.10 | 34.00 | 14.64 | 14.67 | 10.70 | 2.67 | 32.13 | 48.67 | 0.01 | 5.42 | 100.00 |
| Arts | 44.16 | 21.43 | 13.27 | 2.38 | 6.95 | 2.38 | 34.05 | 73.81 | 0.00 | 1.57 | 100.00 |
| Biochem Mol Biol | 33.71 | 25.83 | 13.92 | 16.89 | 17.55 | 18.21 | 29.89 | 39.07 | 0.00 | 4.93 | 100.00 |
| Business | 46.65 | NULL | 16.35 | NULL | 12.04 | NULL | 23.19 | NULL | 0.04 | 1.73 | 100.00 |
| Chem Eng | 34.04 | NULL | 31.42 | NULL | 8.65 | NULL | 15.90 | NULL | 0.01 | 9.98 | 100.00 |
| Chemistry | 36.09 | 28.57 | 19.73 | 14.29 | 11.00 | 14.29 | 23.06 | 42.86 | 0.00 | 10.12 | 100.00 |
| Computer | 29.35 | 0.00 | 37.53 | 100.00 | 10.28 | 0.00 | 18.91 | 0.00 | 0.02 | 3.91 | 100.00 |
| Decision | 17.46 | NULL | 57.79 | NULL | 15.79 | NULL | 7.04 | NULL | 0.00 | 1.92 | 100.00 |
| Dentistry | 41.37 | NULL | 8.86 | NULL | 9.03 | NULL | 38.94 | NULL | 0.00 | 1.80 | 100.00 |
| Earth | 39.94 | 42.86 | 17.40 | 14.29 | 3.75 | 0.00 | 33.00 | 42.86 | 0.06 | 5.85 | 100.00 |
| Economics | 48.02 | 50.00 | 20.13 | 50.00 | 8.03 | 0.00 | 22.33 | 0.00 | 0.36 | 1.13 | 100.00 |
| Energy | 58.57 | NULL | 12.21 | NULL | 3.58 | NULL | 9.49 | NULL | 0.00 | 16.15 | 100.00 |
| Engineering | 42.39 | 0.00 | 19.35 | 100.00 | 7.23 | 0.00 | 25.17 | 0.00 | 0.03 | 5.83 | 100.00 |
| Environmental | 42.61 | 40.98 | 15.26 | 19.67 | 3.11 | 0.00 | 34.04 | 39.34 | 0.01 | 4.97 | 100.00 |
| Health | 36.42 | NULL | 16.03 | NULL | 10.36 | NULL | 35.90 | NULL | 0.02 | 1.27 | 100.00 |
| Immunology | 35.33 | 31.25 | 11.63 | 9.38 | 15.86 | 4.69 | 31.02 | 54.69 | 0.00 | 6.16 | 100.00 |
| Materials | 41.42 | NULL | 15.56 | NULL | 9.30 | NULL | 24.19 | 0.00 | NULL | 9.53 | 100.00 |
| Mathematics | 24.31 | 27.03 | 45.07 | 32.43 | 13.58 | 13.51 | 12.57 | 27.03 | 0.04 | 4.43 | 100.00 |
| Medicine | 37.87 | 29.67 | 10.94 | 11.39 | 10.51 | 22.78 | 39.19 | 36.16 | 0.00 | 1.49 | 100.00 |
| Multidisciplinary | 37.84 | 38.38 | 11.33 | 11.11 | 15.24 | 16.16 | 31.26 | 34.34 | 0.00 | 4.33 | 100.00 |
| Neuroscience | 34.10 | 40.98 | 11.49 | 11.48 | 13.35 | 3.28 | 39.64 | 44.26 | 0.01 | 1.41 | 100.00 |
| Nursing | 37.27 | 40.00 | 10.47 | 20.00 | 8.97 | 0.00 | 42.23 | 40.00 | 0.00 | 1.06 | 100.00 |
| Pharmacology | 38.78 | 57.14 | 9.97 | 3.57 | 10.45 | 3.57 | 36.73 | 35.71 | 0.00 | 4.08 | 100.00 |
| Physics | 35.76 | 60.00 | 25.19 | 20.00 | 11.16 | 0.00 | 21.13 | 20.00 | 0.05 | 6.71 | 100.00 |
| Psychology | 43.64 | 37.93 | 13.05 | 6.90 | 5.65 | 0.00 | 37.18 | 55.17 | 0.00 | 0.48 | 100.00 |
| Social | 36.22 | 54.41 | 12.91 | 7.35 | 18.59 | 2.94 | 29.29 | 35.29 | 0.00 | 2.99 | 100.00 |
| Veterinary | 40.76 | 63.41 | 13.30 | 26.83 | 6.37 | 2.44 | 37.06 | 7.32 | 0.00 | 2.51 | 100.00 |

Note: **A** means after data processing, and **B** means before data processing.

To better understand the importance of our work that expanding the journal list, we compare the distribution of $P_{ij}$ (percent of the number of citations in total) before and after the journal expansion. We obtain Table 5 that shows $P_{ij}$ after and before the expansion of the journal list. In Table 5, we find that before data processing, there are only 20 disciplines, and many values are 0. We obtain no citation records in section MR and section RD, and so the results only include four section types: I, M, R and D. Therefore, we just compare the distributions of $P_{ij}$ from 20 disciplines in four section types before and after the expansion work in Fig. 9.

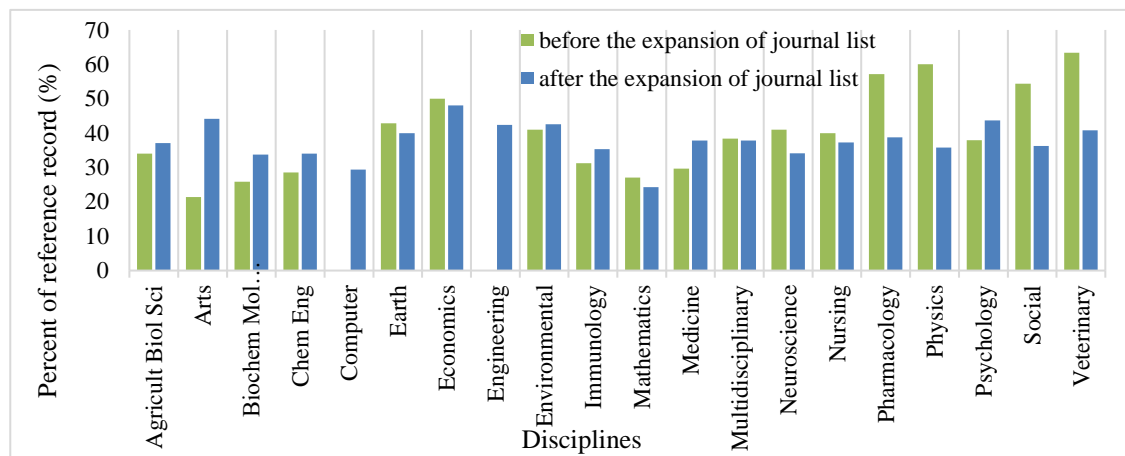

(a) Percent of citations in section I of all disciplines before and after the expansion of journal list

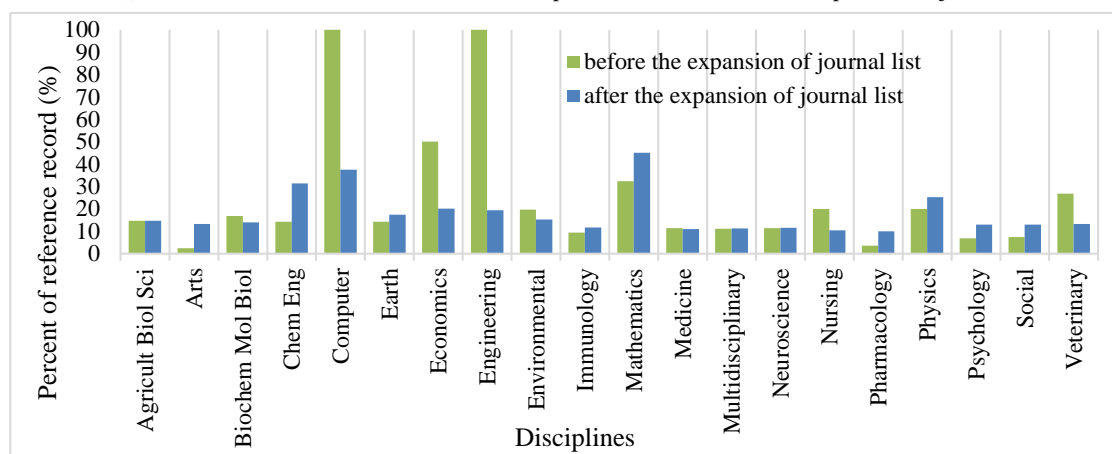

(b) Percent of citations in section M of all disciplines before and after the expansion of journal list

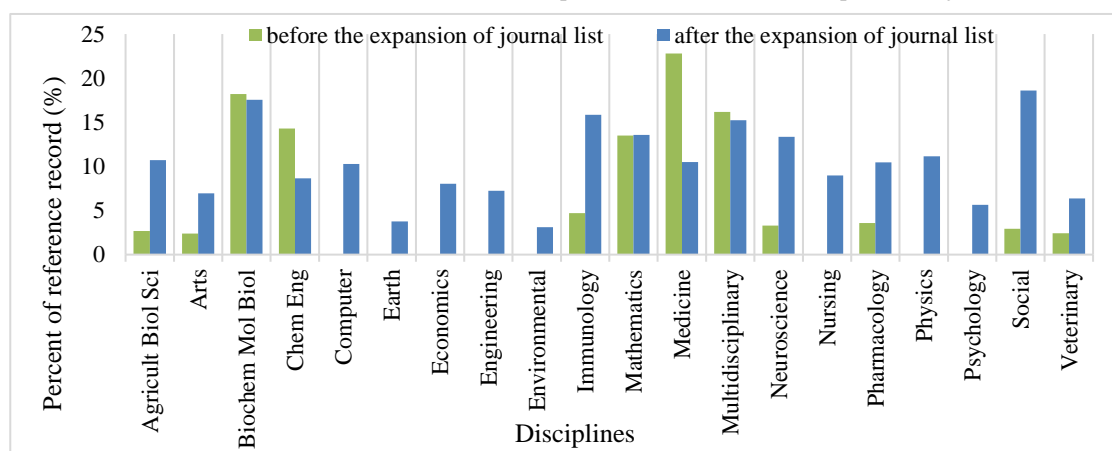

(c) Percent of citations in section R of all disciplines before and after the expansion of journal list

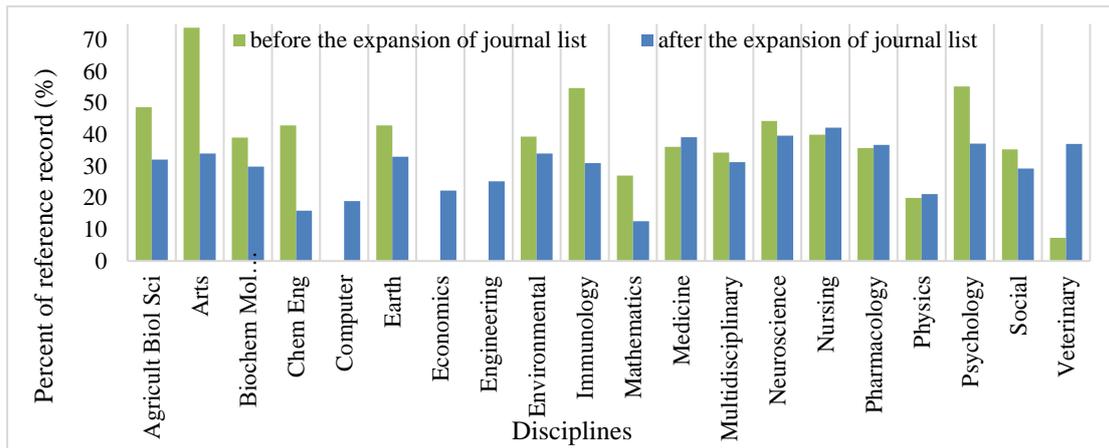

(d) Percent of citations in section D of all disciplines before and after the expansion of journal list

**Fig. 9.** Percent of citations in each section type of all disciplines before and after the expansion of journal list

Some disciplines in Table 5, including Computer, Economics and Engineering, are abnormal. Especially, Computer and Engineering have only been cited in one section type. For other disciplines, we find that the distributions of Medicine, Multidisciplinary and Neuroscience are similar to the results after the expansion of the journal list while the others are different. In addition, as shown in Fig. 9, before journal expansion, the distributions of Computer and Engineering are the same in Fig. 9(b), which is completely the opposite of our final result. In addition, the distributions in sections M, R and D are significantly different. Many disciplines almost appear in one section before journal expansion, such as Arts that mostly appears in section D (Fig. 9(d)), and its proportions in section I (Fig. 9(a)) and M (Fig. 9(b)) are less than the value after journal expansion. In addition, some disciplines are just the opposite. Mathematics appears more after the expansion of journal in section M, as shown in Fig. 9(b), and less in other section types. The reason is the size of the data. Before the expansion of the journal list, we only obtained limited records with disciplines, and so the results are incomplete and untrustworthy. If we analyze the original data directly and without the expansion of the journal list, the results we obtain will be quite different.

The comparison of the distributions of different disciplines before and after the expansion of the journal list demonstrates that identifying the disciplines of more references is important. It is essential to characterize references from different disciplines.

## 6 Conclusion and future works

In this study, we characterize the references from different disciplines based on 210,334 full-text articles on PLoS. Regarding the data processing, our essential work is expanding the journal list with specific disciplines. There are six section types (I, M, R, D, MR and RD) in this paper to show the cited positions. Based on the discipline, location and content of citations information, we characterize the differences and test the significance of the references regarding the number of citations, average cited intensity and average citation length. Then, we use citation analysis and CCA to obtain the distributions of the references from different disciplines.

Regarding the number of citations, Fig. 4 shows that basic disciplines such as Mathematics also provide large proportions of citations, and Social is also important in many research fields such as life science. We combine the number of citations, citation position and citation content of each discipline. Fig. 6 shows the distributions of the 27 disciplines in different section types are different.

In addition, the results also indicate that there are both practical and tested significant differences in the distributions in sections I, M, R and D. Computer, Decision and Mathematics provide most of the research methods. This indicates that basic disciplines are also considered to be important components, and the concerns of the articles and are valued by researchers. Disciplines such as Medicine and Nursing relevant to the fields of PLoS play important roles in section D. We find that both disciplines in section R are close to the field of the dataset, and basic disciplines are also important in the results.

We finally compare the distributions of citations before and after data processing, and we identify the disciplines of references. The differences in the distributions between these two situations are shown in Figs. 8 and 9. The results are quite different without this work, and the conclusion is incomplete and untrustworthy.

Our study focuses on section-level and sentence-level information. We use traditional citation analysis and CCA to characterize the references from different disciplines. Our work shows that ignoring important data will obtain different results, or the results will be faulty. The dataset and results in this research are essential to detecting the resources of referred information and exploring the contributions of different disciplines in research works. Our work can help to identify the knowledge that from references flows into the research papers. In future work, we will consider the two sides communication process, and focus on the specific interaction among disciplines and improve our approach. We will extract and analyze more features in citation content using entity extraction and reveal the details of the interdisciplinary knowledge flow in different research fields and in different citation locations.

## Acknowledgments

This work is supported by National Natural Science Foundation of China (Grant No.72074113) and Science Fund for Creative Research Group of the National Natural Science Foundation of China (Grant No. 71921002).

## Appendix A: Full name and abbreviations of 27 disciplines

| Id | Reference discipline | Reference discipline (abbreviated form) |
|---|---|---|
| 1 | Medicine | Medicine |
| 2 | Biochemistry Genetics and Molecular Biology | Biochem Mol Biol |
| 3 | Agricultural and Biological Sciences | Agricult Biol Sci |
| 4 | Multidisciplinary | Multidisciplinary |
| 5 | Neuroscience | Neuroscience |
| 6 | Mathematics | Mathematics |
| 7 | Immunology and Microbiology | Immunology |
| 8 | Social Sciences | Social |
| 9 | Psychology | Psychology |
| 10 | Environmental Science | Environmental |
| 11 | Pharmacology Toxicology and Pharmaceutics | Pharmacology |
| 12 | Earth and Planetary Sciences | Earth |
| 13 | Veterinary | Veterinary |
| 14 | Arts and Humanities | Arts |
| 15 | Chemistry | Chemistry |
| 16 | Nursing | Nursing |
| 17 | Engineering | Engineering |
| 18 | Computer Science | Computer |
| 19 | Physics and Astronomy | Physics |
| 20 | Materials Science | Materials |
| 21 | Economics Econometrics and Finance | Economics |
| 22 | Health Professions | Health |
| 23 | Business Management and Accounting | Business |
| 24 | Dentistry | Dentistry |
| 25 | Decision Sciences | Decision |
| 26 | Chemical Engineering | Chem Eng |
| 27 | Energy | Energy |

**Appendix B: Average cited intensity of 27 disciplines in 6 section types**

| Discipline | Section I | Section M | Section R | Section D | Section MR | Section RD | $ACI_i$ | Standard Deviations |
|---|---|---|---|---|---|---|---|---|
| Agricult Biol Sci | 1.27 | 1.29 | 1.45 | 1.34 | 1.47 | 1.45 | 1.32 | 0.08909 |
| Arts | 1.31 | 1.30 | 1.48 | 1.29 | 1.50 | 1.48 | 1.32 | 0.1027 |
| Biochem Mol Biol | 1.24 | 1.25 | 1.38 | 1.27 | 1.40 | 1.43 | 1.28 | 0.08424 |
| Business | 1.28 | 1.35 | 1.81 | 1.27 | 1.00 | 1.44 | 1.34 | 0.26574 |
| Chem Eng | 1.19 | 1.24 | 1.27 | 1.22 | 1.00 | 1.41 | 1.24 | 0.13288 |
| Chemistry | 1.21 | 1.20 | 1.33 | 1.25 | 0.00 | 1.38 | 1.25 | 0.5248 |
| Computer | 1.23 | 1.25 | 1.32 | 1.27 | 1.00 | 1.31 | 1.26 | 0.11781 |
| Decision | 1.16 | 1.15 | 1.38 | 1.13 | 0.00 | 1.38 | 1.19 | 0.51906 |
| Dentistry | 1.19 | 1.28 | 1.96 | 1.31 | 1.00 | 1.34 | 1.29 | 0.32457 |
| Earth | 1.28 | 1.28 | 1.39 | 1.38 | 1.60 | 1.50 | 1.33 | 0.12582 |
| Economics | 1.24 | 1.28 | 1.44 | 1.25 | 1.33 | 1.52 | 1.27 | 0.11325 |
| Energy | 1.18 | 1.21 | 1.12 | 1.15 | 0.00 | 1.20 | 1.18 | 0.47961 |
| Engineering | 1.21 | 1.32 | 1.34 | 1.27 | 1.25 | 1.34 | 1.26 | 0.05345 |
| Environmental | 1.26 | 1.33 | 1.33 | 1.33 | 1.44 | 1.43 | 1.31 | 0.0689 |
| Health | 1.29 | 1.25 | 2.17 | 1.35 | 1.25 | 1.44 | 1.36 | 0.356 |
| Immunology | 1.26 | 1.33 | 1.42 | 1.30 | 1.60 | 1.45 | 1.32 | 0.1242 |
| Materials | 1.24 | 1.26 | 1.33 | 1.28 | 1.00 | 1.39 | 1.28 | 0.13387 |
| Mathematics | 1.26 | 1.18 | 1.36 | 1.25 | 1.75 | 1.38 | 1.24 | 0.20344 |
| Medicine | 1.20 | 1.28 | 1.54 | 1.28 | 1.74 | 1.34 | 1.27 | 0.20373 |
| Multidisciplinary | 1.26 | 1.36 | 1.47 | 1.32 | 1.51 | 1.53 | 1.33 | 0.11053 |
| Neuroscience | 1.25 | 1.34 | 1.43 | 1.31 | 1.72 | 1.41 | 1.31 | 0.16553 |
| Nursing | 1.20 | 1.25 | 2.13 | 1.28 | 1.00 | 1.30 | 1.29 | 0.39248 |
| Pharmacology | 1.20 | 1.33 | 1.34 | 1.25 | 0.00 | 1.35 | 1.25 | 0.53154 |
| Physics | 1.22 | 1.23 | 1.40 | 1.27 | 1.56 | 1.41 | 1.26 | 0.13258 |
| Psychology | 1.31 | 1.26 | 1.67 | 1.30 | 1.20 | 1.26 | 1.32 | 0.16943 |
| Social | 1.29 | 1.24 | 1.50 | 1.31 | 1.00 | 1.45 | 1.33 | 0.1768 |
| Veterinary | 1.23 | 1.32 | 1.33 | 1.32 | 1.00 | 1.46 | 1.28 | 0.15423 |
| Average of all records | 1.23 | 1.28 | 1.44 | 1.29 | 1.54 | 1.43 | 1.29 | 0.11957 |
| Standard Deviations | 0.00781 | 0.01019 | 0.04868 | 0.01033 | 0.10413 | 0.01493 | 0.00843 | |

**Appendix C: Average citation length of 27 disciplines in 6 section types**

| Discipline | Section I (words) | Section M (words) | Section R (words) | Section D (words) | Section MR (words) | Section RD (words) | ACL$_i$ (words) | Standard Deviations |
|---|---|---|---|---|---|---|---|---|
| Agricult Biol Sci | 30.33 | 29.95 | 31.83 | 31.03 | 28.71 | 31.19 | 30.71 | 1.10101 |
| Arts | 31.35 | 31.48 | 33.04 | 31.78 | 23.67 | 31.50 | 31.63 | 3.3887 |
| Biochem Mol Biol | 30.04 | 29.30 | 31.46 | 30.45 | 30.16 | 31.45 | 30.38 | 0.84741 |
| Business | 30.17 | 30.27 | 31.73 | 30.15 | 17.00 | 32.64 | 30.41 | 5.80195 |
| Chem Eng | 31.54 | 25.26 | 31.86 | 31.30 | 26.00 | 32.89 | 29.69 | 3.29 |
| Chemistry | 31.09 | 28.50 | 32.09 | 30.68 | 0.00 | 32.68 | 30.75 | 12.74062 |
| Computer | 30.09 | 28.83 | 31.21 | 31.40 | 27.00 | 31.46 | 30.03 | 1.78509 |
| Decision | 28.85 | 29.11 | 33.33 | 29.56 | 0.00 | 36.18 | 29.90 | 13.1427 |
| Dentistry | 29.80 | 31.40 | 30.34 | 29.96 | 30.00 | 29.93 | 30.06 | 0.59687 |
| Earth | 31.40 | 32.40 | 34.55 | 32.91 | 24.63 | 34.49 | 32.37 | 3.68592 |
| Economics | 30.82 | 30.04 | 30.08 | 31.33 | 40.38 | 34.24 | 30.79 | 4.01562 |
| Energy | 30.65 | 36.11 | 32.10 | 33.44 | 0.00 | 36.06 | 32.51 | 13.91503 |
| Engineering | 31.25 | 30.41 | 31.50 | 30.81 | 36.40 | 33.00 | 31.10 | 2.22728 |
| Environmental | 30.52 | 30.75 | 32.88 | 31.54 | 33.69 | 31.18 | 31.01 | 1.25871 |
| Health | 31.30 | 31.94 | 32.96 | 32.23 | 38.60 | 28.98 | 31.88 | 3.20838 |
| Immunology | 30.07 | 29.18 | 31.68 | 30.99 | 29.13 | 31.54 | 30.60 | 1.13932 |
| Materials | 31.02 | 28.11 | 30.90 | 30.34 | 19.00 | 32.77 | 30.56 | 4.97824 |
| Mathematics | 30.53 | 28.24 | 31.59 | 31.08 | 36.44 | 30.53 | 29.71 | 2.72175 |
| Medicine | 31.08 | 31.02 | 32.63 | 31.61 | 32.70 | 31.80 | 31.45 | 0.72959 |
| Multidisciplinary | 30.72 | 30.28 | 32.13 | 31.27 | 31.04 | 32.10 | 31.12 | 0.74336 |
| Neuroscience | 30.52 | 30.51 | 31.94 | 30.98 | 25.27 | 31.20 | 30.90 | 2.4099 |
| Nursing | 31.63 | 32.12 | 34.56 | 32.06 | 46.00 | 30.39 | 32.11 | 5.81362 |
| Pharmacology | 30.23 | 30.28 | 32.09 | 30.33 | 0.00 | 31.57 | 30.52 | 12.63881 |
| Physics | 31.63 | 30.78 | 32.05 | 31.58 | 48.12 | 32.99 | 31.55 | 6.69903 |
| Psychology | 31.37 | 31.24 | 34.44 | 31.95 | 36.00 | 31.44 | 31.74 | 1.99793 |
| Social | 30.76 | 32.86 | 32.32 | 31.51 | 38.00 | 31.41 | 31.56 | 2.64693 |
| Veterinary | 30.24 | 29.94 | 32.39 | 31.58 | 34.50 | 32.05 | 30.88 | 1.6517 |
| Average of all records | 30.60 | 30.05 | 31.98 | 31.15 | 31.70 | 31.59 | 30.93 | 0.73385 |
| Standard Deviations | 0.12665 | 0.37528 | 0.21441 | 0.1656 | 2.60851 | 0.31802 | 0.15028 | |